\documentclass[10pt,preprint]{sigplanconf}
\usepackage{times}

\usepackage{datetime}
\usepackage{url}
\usepackage{hyperref}
\usepackage{subcaption}
\hypersetup{
  colorlinks,
  linkcolor={red},
  citecolor={blue},
  urlcolor={blue}
}

\conferenceinfo{SOSP'15}{October 4--7, 2015, Monterey, CA}
\copyrightyear{2015} 

\date{}

\authorinfo{Piyus Kedia (Microsoft Research) and Sorav Bansal (IIT Delhi)}{}

\begin{document}

\title{A Software-only Mechanism for Device Passthrough and Sharing} 
\maketitle

\begin{abstract}

Network processing elements in virtual machines, also known
as Network Function Virtualization (NFV) often face CPU bottlenecks
at the virtualization interface. Even highly optimized paravirtual
device interfaces fall short of the throughput requirements of modern
devices. Passthrough devices, together with SR-IOV support for
multiple device
virtual functions (VF) and IOMMU support, mitigate this problem
somewhat, by
allowing a VM to directly control a device partition
bypassing the virtualization stack. However, device passthrough is
riddled with its own problems of low consolidation ratios, relatively static
resource partitioning, and difficulties in VM migration.

We present a paravirtual interface that securely exposes an I/O device directly
to the guest OS running inside the VM, and yet allows that device
to be securely shared
among multiple VMs and the host.
Compared to the best-known paravirtualization interfaces, our
paravirtual interface
supports up to 2x higher throughput, and is closer in performance
to device passthrough. Unlike device passthrough however, we do not require
SR-IOV or IOMMU support, and allow
fine-grained dynamic resource allocation, significantly higher
consolidation ratios, and seamless VM migration.
Our security mechanism is based on a novel approach called
{\em dynamic binary opcode subtraction}.

\end{abstract}

\section{Introduction}
\label{sec:intro}

Today's networks rely on ``middleboxes'' \cite{honda11, sherry12}
(also called network appliances) for a variety of network processing
needs, such as overlay network switches, firewalls, load balancers,
routers, etc. Network function virtualization (NFV) proposes shifting
middlebox processing from specialized
hardware appliances to software running
on commodity hardware. Further, NFV benefits significantly
from virtualization capabilities, to significantly improve resource
utilization \cite{ghodsi12}, and allow sharing of hardware
resources among multiple
(and potentially untrusted) tenants \cite{clickos, ghodsi12}.

Virtualized network appliances, however present a new set of performance
challenges.
For example, ClickOS \cite{clickos} showed that the network stack on
the Xen hypervisor \cite{xen} falls far short of the maximum
achievable throughput on a 10Gbps NIC, using commodity x86 hardware.
Because the original network stacks were
not designed for such high-throughput workloads,
inefficiencies lurk at multiple levels in current
network stacks: (a) guest-side and host-side user/kernel network API
(e.g., socket API) was not designed to handle
such workloads;  (b) the device virtualization interface between
the guest and the host
(e.g., virtio \cite{virtio}) is often a performance
bottleneck; and (c) the host-side network bridge/switch (e.g., Linux
bridge, Open vSwitch \cite{openvswitch}) is usually incapable of handling high
rates of traffic.

The {\tt netmap} framework \cite{netmap}
proposes an efficient user/kernel interface, best
suited for high-throughput I/O. A netmap-capable
user process maps shared-memory producer-consumer rings to
communicate efficiently with the kernel.
This allows a zero-copy interface between the user and the
kernel, and also allows the user to perform I/O in batches,
thus amortizing the cost of traversing the kernel's network stack
over multiple packets. This results in high overall throughputs.
For example,
netmap's {\tt pkt-gen} \cite{netmap} can saturate a 10Gbps link
with a single 3GHz core
using 64B packets (14.88 Mpps), while the socket-based network
stack reaches only about
a third of the maximum achievable throughput \cite{netmap}.

However, if {\tt netmap} is used inside
a VM, performance bottlenecks emerge at the device virtualization
interface. Software-emulated devices exhibit significant CPU overheads
related to faithful execution of the device state machine in software.
Paravirtual interfaces for device virtualization
are relatively faster, but even they prove inadequate in such
high-throughput settings. Table~\ref{tab:thruput} shows the throughput
of running netmap's {\tt pkt-gen} inside a virtual machine using
the {\tt virtio}-based paravirtual device \cite{virtio}.
The virtio
interface involves shared-memory communication between the
guest and the host through producer-consumer rings.
The
table shows results for three different implementations of the host-side
networking stack, namely {\tt tap}, {\tt tap-vhost}, and {\tt netmap}
(more details in Section~\ref{sec:experiments}).
Compared to bare-metal, the performance penalty of the {\tt virtio}
interface is significant. The penalty is even more pronounced in
CPU-constrained settings. To show this, we also show experiments with
uniprocessor hosts using the {\tt virtio} interface. While guest-host
communication involves cacheline transfers in multiprocessor hosts,
uniprocessor hosts require expensive VM exits.
Further, sharing the network device among multiple VMs incurs
performance penalties at the host-side switch.

These CPU bottlenecks indicate that our I/O virtualization stacks
are perhaps a bit too ``deep''. The cost of traversing the I/O
virtualization stack (even with paravirtualization)
is often more than the cost of actual network
processing. This I/O virtualization cost stems primarily from the need
to secure the host and the device from untrusted VMs, which forces
us to use narrow interfaces between the VM and the host (such as
producer-consumer rings) resulting
in relatively deep I/O stacks.

\begin{table}
\begin{center}
\begin{subtable}{0.5\textwidth}
\begin{tabular}{c|c|c|c|c}
     & \multicolumn{2}{|c|}{tx(Kpps)} & \multicolumn{2}{|c}{rx(Kpps)} \\
    \cline{2-5}
		& 60B & 1500B & 60B & 1500B\\
    \cline{2-5}
     socket-baremetal & 470 & 180 & 394 & 214 \\
    \hline
     socket-virtio-vhost & 250 & 170 & 300 & 150\\
    \hline
     netmap-baremetal & 14810 & 820 & 13304 & 820 \\
    \hline
     netmap-virtio & 236 & 193 & 306 &  268 \\
    \hline
     netmap-virtio-vhost & 357 & 285 & 416 & 422 \\
    \hline
     netmap-virtio-netmap & 154 & 154 & - & -\\
    \hline
     netmap-fastio & {\bf 14704} & {\bf 815} &  {\bf 13292} & {\bf 816} \\
    \hline
     netmap-fastio-no-rzc & - & - & 12970 & 813 \\
\end{tabular}
\subcaption{\label{tab:thruput}}
\end{subtable}
\end{center}

\begin{center}
\begin{subtable}{0.5\textwidth}
\begin{tabular}{c|c|c|c|c}
     & \multicolumn{2}{|c|}{tx(Kpps)} & \multicolumn{2}{|c}{rx(Kpps)} \\
    \cline{2-5}
		& 60B & 1500B & 60B & 1500B \\
    \cline{2-5}

     netmap-baremetal & 14600 & 822 &  13620 &  820 \\
    \hline
     netmap-virtio &  188 &  182 &  35 &  20 \\
    \hline
     netmap-virtio-vhost & 331 & 256 & 68 & 41 \\
    \hline
     netmap-fastio & 14632 & 815 & 13001 & 816 \\
    \hline
     netmap-fastio-no-rzc & - & - & 13049 & 811 \\
\end{tabular}
\subcaption{\label{tab:thruput_uni}}
\end{subtable}
\caption{Single-guest throughput on (a) multiprocessor
and (b) uniprocessor hosts with a 10Gbps NIC. All numbers are given
in Kpps (thousand packets per second). To calculate
throughput, use Mbps = Kpps * (pktsize + 20.2) * 8.
{\tt netmap-fastio-no-rzc} refers to fastio without receive-side zero-copy}
\end{center}
\end{table}

In this paper, we show that security can be provided in alternate
ways; we use this observation to make the I/O stack significantly thinner.
We allow a VM to have direct visibility into the hardware device. The
VM can read/write to the hardware device without host intervention.
Yet, we ensure that an untrusted VM cannot harm the
host and/or other guests.

One of the central ideas in this paper is
{\em dynamic binary opcode subtraction}, or DBOS.
DBOS enables the hypervisor to restrict VM behaviour;
we use DBOS to implement the requisite security required
in the I/O virtualization stack.
Table~\ref{tab:thruput} shows results using our approach
(labeled {\tt netmap-fastio}) to paravirtualization. Using
{\tt fastio}, the I/O throughput
achievable inside the VM is comparable to bare-metal performance, even
on a single processor.
We achieve this for an off-the-shelf guest OS (Linux), without
assumptions about
SR-IOV and IOMMU hardware support. Our interface supports fair allocation
of device
resources among untrusted VMs, and allows fast switching among them.
Compared to netmap-based VALE \cite{vale}, our software switch provides
up to 10x higher throughput.

In summary, this paper makes two primary contributions:
\begin{itemize}
\item We present a fast
device paravirtualization mechanism which exhibits close
to bare-metal performance. Compared to conventional
paravirtualization (e.g., {\tt virtio}),
our scheme provides up to 25x higher throughput. Compared to highly
hand-optimized I/O virtualization stacks (e.g., ClickOS \cite{clickos}), we
achieve around
2x throughput improvement for small 60B packets, and around 30\% throughput
improvement for 1500B packets.
\item We introduce a novel security mechanism, DBOS,
which allows a hypervisor to restrict guest behaviour. We demonstrate
an application of DBOS to improve I/O virtualization performance.
\end{itemize}

This paper is organized as follows. Section~\ref{sec:background} provides
relevant background on network processing and switching; we discuss our
DBOS-based solution
alongwith its security considerations in Section~\ref{sec:solution}. The operation
of our guest-side paravirtual driver is presented in Section~\ref{sec:fastio},
Section~\ref{sec:experiments} presents our experiments and results;
Section~\ref{sec:discussion} discusses design considerations and alternate design
choices; Section~\ref{sec:relwork} discusses related work, and
Section~\ref{sec:conclusion} concludes.

\section{Background}
\label{sec:background}
Previous work \cite{netmap} has shown that the traditional user/kernel interface for
network processing can become a performance bottleneck for high-throughput workloads.
The netmap API defines a user/kernel interface, whereby a user process can pre-allocate
a set of ring buffers to communicate with the kernel, and map this allocated memory
in its address space. The interface between the user and the kernel is that of a
shared ring containing buffer pointers. On the transmit path, the user can write
the packet contents in pre-allocated buffers, setup the buffer pointers in the
shared ring, and increment the ring's head pointer. The kernel would consume
buffers from the shared ring by incrementing the tail pointer, and
send them to the network port/device. Similarly, on the receive path, the user
would first set up empty
buffers in its ring and update the head pointer; the kernel would copy received
packets to the ring buffers, and increment the ring's tail pointer (so the user
can read the packet contents).
Unlike traditional socket API, netmap does not involve overheads
related to memory allocation/deallocation, copying, and other book-keeping
on the I/O datapath. The overall result is a much faster user/kernel API. As we
see in Table~\ref{tab:thruput}, the netmap API can saturate a 10Gbps NIC
(14.8 Mpps), while the
traditional socket API can only reach a fraction of the link capacity for
60B packets on baremetal.

Running the network processing elements inside a virtual machine require
device virtualization. Unlike full virtualization, paravirtualization allows
flexibility in choosing the right VM/hypervisor interface for optimum performance.
Today, device paravirtualization is typically done using a shared-memory producer-consumer
ring between the guest and the host. For example, KVM/Virtio sets up
shared-memory rings between the guest and the Qemu process.
On the transmit
path, the guest writes to the shared ring and the host-side Qemu process reads
from it. Similarly, data flows in the reverse direction on the receive path.
As an optimization, the host-side kernel may read/write to the queue (virtio-vhost)
instead of the Qemu process.


A host-side software switch (e.g., Linux tun/tap, Open vSwitch), typically
implemented as a part of the kernel multiplexes/demultiplexes packets for
multiple guests. Current switches are unable to sustain
high throughputs; e.g., the tap interface on Linux and Open vSwitch
peak at around 300-600Kpps \cite{vale, clickos}.
A recent software-switch,
VALE, takes advantage of the fact that its ports may be using the netmap
API. Using this, VALE implements switching in batches, thus exposing
opportunities for improving forwarding performance, and optimizing
cache utilization through prefetching. Even with these optimizations, a VALE switch
together with the virtio-vhost interface, can handle only up to 3.5Mpps \cite{vale}.

Due to these limitations, a common approach today for high-performance
networking with virtual machines, is device-passthrough, whereby a NIC can
be exposed directly to a VM. Device passthrough reduces scalability, as
the device is exclusively controlled by the given virtual machine. This
problem is mitigated by modern NICs supporting hardware multi-queuing,
VMDq, and SR-IOV \cite{intel_network_support}. Further, device passthrough complicates
live migration, and requires IOMMU support for security.

We show that it is possible to achieve equivalent performance as device-passthrough
without compromising scalability, or live-migration. We also do not require
SR-IOV or IOMMU hardware support.

\section{Our Solution}
\label{sec:solution}
Our paravirtual device driver, called {\tt fastio} driver,
is different from current paravirtual drivers
(e.g., virtio) in several ways. Firstly, we require that our device driver
should authenticate itself with the hypervisor at load-time. Once authenticated,
the hypervisor can trust the fastio driver with privileged device
state. The authentication and trust is maintained in multiple steps:
\begin{enumerate}
\item Our fastio driver does not rely on any read/write static data, i.e., its
binary object file contains only code pages and read-only static data. For all
other memory needs, the driver
must use ``special'' stack space or
heap memory. As we see later, the hypervisor
ensures that this stack and heap memory remain private to our driver, i.e., the
rest of the guest kernel cannot read/write to it.
\item At load time, the fastio driver loads its code and read-only data pages in the
guest memory, and informs the hypervisor about its loaded addresses and size, using
a hypercall.
\item Using page-protection bits in the x86 extended page
tables (EPTs) \cite{intelRefManual3B}, the
hypervisor write-protects all code and data pages of the fastio driver. This allows
the hypervisor to ensure that the guest OS cannot change the driver code/data, after
it has been loaded and authenticated.
\item The hypervisor verifies the contents of the driver's code pages. We
perform this verification using
cryptography-based digital-signatures. The fastio driver
presents a signed certificate (signed by the hypervisor) which
certifies the contents of the code pages; the hypervisor
computes a sha1 hash of the code
pages, and ensure that it matches with the
value presented in the signed certificate.
\end{enumerate}


Next, we expose the privileged device state to the guest, by mapping its memory
addresses (including MMIO addresses) in the guest's physical address space (GPA
space). This involves creating mappings in the guest's EPT for the device data
structures at ``privileged GPA addresses'', or PGPAs. We need
to ensure that the PGPAs are distinct from the
actual guest physical memory addresses, to avoid conflicts. In our
prototype implementation, we use addresses above
4GB for PGPAs; we assume that all our guests have less than 4GB RAM.
This is not a fundamental restriction, as PGPAs can be chosen to be arbitrarily
large in a 64-bit address space.
Mapping privileged state at PGPAs, exposes the hypervisor to
attacks from the untrusted guest. To prevent these attacks, we next ensure that
the PGPAs cannot be mapped in the guest's virtual address space, unless the
hypervisor explicitly allows it to do so. Essentially, we will try to
ensure that the fastio driver is allowed to map the PGPAs in its address space,
while the rest of the guest kernel is not allowed to do so.

If the hypervisor can successfully ensure that the
PGPAs are not mapped in the guest's virtual address space (GVA space), it
effectively ensures that the guest cannot access the PGPA addresses.
We configure the virtualization hardware to ensure
that a VM exit occurs on every change to the guest's virtual address space. i.e.,
an exit should occur on every execution of the {\tt mov-to-cr3}, {\tt mov-to-cr4},
{\tt mov-to-cr0}, and other privileged instruction that can
potentially change the VA space. On VM exits resulting
from these instructions, the hypervisor checks the
new GVA space to ensure that no mappings exist to our PGPAs in it. For example,
on the execution of a {\tt mov-to-cr3} instruction inside the guest, a VM
exit occurs, and
the hypervisor walks the page table to ensure that none of
its entries point to the PGPAs.

Because x86 paging allows changes to the VA space through simple modifications
to the page-table entries, we further mark all the page-table pages read-only
on every cr3 load.
Thus, while walking the guest's page table during the VM-exit caused by the
{\tt mov-to-cr3} instruction, the hypervisor marks all the GPAs corresponding
to the page-table pages as read-only. Any future write access by the guest
to its page-table pages, causes an EPT violation, resulting in a VM exit.
The hypervisor then emulates the exiting instruction within the hypervisor
before returning control back to the VM.

Using this, we ensure that the guest can never directly access the
PGPA space. This solution requires VM exits on every execution of
the {\tt mov-to-cr3} instruction
(among other such instructions that can change the address space) and on
every write access to a page-table page within the guest. Now, the hypervisor
needs to implement a mechanism which allows our fastio driver to access the
PGPA space directly (but still disallows the rest of the guest kernel from doing so).

The hypervisor sets up
a special page-table, called the privileged page table {\em PPT}, using pages
in the read-only data section of the fastio driver. Notice that the hypervisor
is free to write to the PPT, even though the PPT pages appear read-only to
the guest kernel. The PPT will contain mappings to the PGPA space, and the fastio
driver can switch to it using the {\tt mov-to-cr3} instruction
to access the device state directly. The PPT would also contain mappings for the
guest's kernel data structures, so that the driver can efficiently communicate
between the device and the guest kernel. In our 32-bit implementation, our PPT
maps the entire guest kernel in the PPT (at addresses above 0xc0000000), and
uses the ``userspace addresses'' (0x0000000-0xc0000000) to map the privileged
device state.

\begin{figure}[htb]
{\small
\begin{verbatim}
fastio_driver() {
   1  save_flags()     # save flags
   2  cli              # clear interrupts
   3  movl $ppt, %eax  # eax <- PPT addr
   4  tmp = read_cr3() # save old cr3 value
   5  movl %eax, %cr3  # load new page table
   6  save_sp()        # save stack pointer
   7  movl $pstk, %esp # use priv stack
   8 
   9  fastio_body()    # calls txsync/rxsync
  10 
  11  restore_sp()     # restore old stackptr
  12  movl tmp, %eax   # eax <- old cr3 value 
  13  movl %eax, %cr3  # load old page table
  14  if (%eax == $ppt)
  15    vmcall         # alert the hypervisor
  16  restore_flags()  # re-enable interrupts
  17  ret              # return to caller
}
\end{verbatim}
}
\caption{\label{fig:fastio_code} Fastio driver pseudo-code}
\end{figure}

The fastio driver's pseudo-code is shown in Figure~\ref{fig:fastio_code}.
The driver first disables interrupts (line 2), then loads the
address of the {\tt PPT} in \%eax register (line 3), and finally,
executes {\tt mov-to-cr3}
to load the PPT (line 5). (We discuss the need for save/restore of the
stack pointer later). The body of the fastio driver can now
access the device
and the guest's data structures to efficiently implement the transmit/receive
logic. In particular, it transfers packets between the guest's netmap ring
and the device's hardware ring.
Finally, after the
body of the fastio driver has executed, the {\tt mov-to-cr3} instruction
is executed to restore the guest's original page table (line 13), before restoring
the original interrupt flag and returning to the caller. (We discuss the
need for lines 14-15 later).

Apart from security considerations, this solution has a serious performance
concern. Every call to the fastio driver involves two executions of the
{\tt mov-to-cr3} instruction, and each of them will cause a VM exit in our
model. The performance overhead of these exits is likely to be more than
the overhead of the {\tt virtio} interface, which only required one exit
(or no exits for multiprocessor hosts). Ideally, we would like to ensure
that the two {\tt mov-to-cr3} instructions executed by our fastio driver
do not cause VM exits, while the other {\tt mov-to-cr3} instructions
executed by the guest kernel cause VM exits.

This differentiation is perhaps hard to achieve efficiently, through runtime
mechanisms alone. We use dynamic binary opcode subtraction (DBOS) to solve this
problem. DBOS involves ensuring that an opcode is not present in the guest's
executable address space. To implement DBOS, the hypervisor removes
execute-privileges from all
guest pages, except the fastio driver's code pages. This is done at the time
when the fastio driver is loaded, and the hypervisor verifies it. Subsequently,
any instruction execution by the guest OS (outside of the fastio driver) would
cause a VM exit resulting from an EPT execute-privileges violation. At this
point, the hypervisor scans the page containing the instruction being
executed to ascertain the absence of the {\tt mov-to-cr3} opcode in that page.
Checking the absence of the {\tt mov-to-cr3} opcode involves checking the
absence of the following byte sequence in the page: {\tt 0x0f}, {\tt 0x20}, ``$B$''.
Here $B$ is any byte that satisfies the equation:
($B \& 0x38 == 0x18$), i.e., the bits 3, 4, and 5 of $B$ should
be equal to 0b011 (cr3). The prime observation is that if a byte sequence corresponding
to an opcode is not even present in the executable address space of a guest, the
guest can never execute that opcode. In the rest
of the paper, we will also call this sequence of bytes representing
the {\tt mov-to-cr3} opcode, the 2.5-byte sequence (two exact bytes, and one byte
with three bits set to a certain value).

We noticed that it is quite rare to find the presence of the 2.5-byte sequence
in typical code. For example, the only code pages in the Linux kernel that
contain this 2.5-byte sequence are the {\tt mov-to-cr3} instructions themselves.
Notice that while ascertaining the absence of a byte sequence, we disregard
any assumptions about instruction boundaries. We call this technique ``dynamic
binary opcode subtraction'', because it {\em subtracts} an opcode dynamically
from the execution stream of a guest.

The {\tt mov-to-cr3} opcode needs to be subtracted not just from the guest
kernel's code stream, but also from the user programs' execution stream
running within the guest. If we disregard user programs, the guest can
launch a simple attack, whereby it can branch to a user code page with
kernel privileges to execute the {\tt mov-to-cr3} instruction. Even in the
user code pages, the presence of the 2.5-byte sequence is extremely rare. In
fact, in all our experiments involving execution of several programs
shipped with stock Ubuntu Linux, including
the SPEC Integer programs, we did not find the presence of the 2.5-byte sequence
in any of them.

If a page containing the current executing instruction (and causing the EPT
violation) does not contain the 2.5-byte sequence, we restore executable
privileges on it. To guard against attacks involving page-boundaries,
we also check the successor and predecessor pages of the currently executing
page. If either of them has already been marked executable, we ensure that
the 2.5-byte sequence does not appear even if the two pages are consider together
as one contiguous block. Similarly, each time an executable page is installed
in a page table (through a page table update, hence causing a VM exit), we
perform the same check again to ensure that the executable page's new neighbours
do not cause the presence of the 2.5-byte sequence.

When we mark a page with execute privileges, we also take away write privileges
from that page (again through manipulation of page-protection bits in the
corresponding EPT entry). If that page is ever written-to
subsequently (to implement
page-swapping, for example), an EPT violation occurs --- in this case,
the hypervisor removes execute privileges from that page, and re-instates write
privileges on that page. This mechanism can also handle dynamically generated
and self-modifying code.

This scheme works well if none of the guest pages contain the 2.5-byte sequence.
However, if a page (or a combination of two successive pages in the GVA space)
indeed contains the 2.5-byte sequence, the hypervisor needs to handle it
gracefully.
A straw-man solution is to
never grant execute privileges to any such page, causing an EPT violation
each time an instruction on that page gets executed. This is likely to result
in a huge slowdown, especially if multiple instructions within a page
execute a large number of times (e.g., a loop). We instead use in-place
binary patching to deal with such situations efficiently.

We patch any instruction containing the 2.5-byte
sequence, with the single-byte {\tt int3} opcode ({\tt 0xcc}), resulting in
a VM exit on its execution. (We configure the virtualization hardware
to generate a VM exit on the {\tt int3} instruction).
The hypervisor keeps track of all such patches, and emulates the original
instruction on the patch-induced VM exits. The use of the {\tt int3} instruction
does not preclude the guest from using it for its own purposes (e.g., debugging),
as the hypervisor can easily differentiate between guest's {\tt int3} and
hypervisor's patched {\tt int3}. The only remaining complication is that
of identifying the instruction boundary containing the 2.5-byte sequence.
As we discussed earlier, in all our experiments, the only occurrence of
the 2.5-byte sequence involved an actual {\tt mov-to-cr3} instruction
within the guest. Hence, simply patching all the 3 bytes
in the sequence would achieve the desired result. If we patch all the 3
bytes, we also take care of cases, where the 2.5-byte sequence straddles two
instructions, i.e., some of the bytes belong to one instruction, while others
belong to the successor instruction.

In general, it is possible that the 2.5-byte sequence appears
in the middle of an
instruction. In this case, if we simply patch the sequence, the guest's
instruction semantics can change (causing the guest to get incorrectly confused).
Here is an example of an instruction that could get incorrectly patched:
\\
\begin{centerline}
{\small
\begin{tabular}{c|c}
assembly & binary representation\\
\hline
mov \$0x18200f, \%eax & 0xb8, 0x0f, 0x20, 0x18, 0x0\\
\end{tabular}
}
\end{centerline}
\\
This instruction contains the 2.5-byte sequence, and if we patched it with
the {\tt int3} opcode, we would replace it with
\\\\
\begin{centerline}
{\small
\begin{tabular}{c|c}
assembly & binary representation\\
\hline
mov \$0xcccccc, \%eax & 0xb8, 0xcc, 0xcc, 0xcc, 0x0\\
\end{tabular}
}
\end{centerline}
\\
Hence, this instruction
would silently behave
incorrectly (without causing a VM exit) if patched by us. To
deal with this situation, we need
to identify the instruction boundary of the instruction containing the 2.5-byte
sequence, and patch its first byte (along with patching the 2.5 bytes sequence
itself). In this example, we should have also replaced the first byte {\tt 0xb8} with
{\tt 0xcc}. Doing so would ensure that a VM exit occurs on the execution of this
instruction, allowing the hypervisor to emulate it correctly.

Our current method for identifying instruction boundaries involves tracking
the values of {\tt eip} for each process/kernel (identified using the
value of the {\tt cr3} register). On noticing a 2.5-byte sequence, we
start disassembling instructions from a known predecessor {\tt eip} in the
current GVA space. Using this disassembly, we can identify the boundary
of the instruction containing the 2.5-byte sequence. If
a predecessor {\tt eip} is not known yet, we simply leave the page without
execute privileges, and emulate the instructions in the
hypervisor if that page executes
again, in the hope that eventually we will find some predecessor {\tt eip}
to the 2.5-byte sequence.
If we still do not find a predecessor {\tt eip} after a large number of
EPT-induced VM exits on that page, we simply patch the 2.5-byte
sequence with the {\tt int3} opcodes.
We had to do this for one page during our experiments on the Linux
kernel, where we found
a {\tt mov-to-cr3} instruction in the first page of the
32-bit Linux kernel image (v3.9.0). None of the instructions preceding this
instruction (including this instruction)
ever executed after loading
the driver, and so we could not reliably determine the instruction
boundary. In this particular case, patching the 2.5-byte sequence
was anyways the correct thing to do. Notice that mis-identification
(or non-identification) of instruction boundaries does not pose a
security risk; it can only cause
misbehaviour within a guest. If a guest is aware of our approach, it can
easily help the VMM by avoiding such situations.

Using DBOS, we configure the virtualization hardware to not
cause VM exits on execution of the {\tt mov-to-cr3} instruction inside
the guest; yet we ensure that the {\tt mov-to-cr3} instruction causes a
VM exit inside the guest kernel, but does not cause an exit
within our fastio driver. This enables us to have an exitless
I/O path between guest-fastio-guest, and yet provide direct visibility
into the hardware device state to fastio. This enables us to
obtain I/O performance close to bare-metal, within the guest. The fastio
driver can now be used, not just to access the hardware device, but also 
to access other privileged state of the host/other VMs, and to implement
fast VM-to-VM and VM-to-host communication.

We next discuss the security threats to our scheme, and our solutions
to them.

\subsection{Security}
\label{sec:security}
Our security model relies on the inability of the guest to change its
virtual address space without hypervisor intervention. We achieve
this by ensuring that the guest's executable address space cannot
contain the {\tt mov-to-cr3} opcode.
We configure the virtualization hardware
such that all guest instructions that can potentially modify
its address space cause VM exits, with the exception of the
{\tt mov-to-cr3} instruction. The {\tt mov-to-cr3} instruction
does not cause a VM exit; instead we use DBOS and binary patching
to ensure that the guest exits on {\tt mov-to-cr3} executions. The
fastio driver's {\tt mov-to-cr3} instructions execute without VM
exits.
We further need to ensure that
the fastio driver's code does not itself contain the 2.5-byte sequence,
except at entry and exit (for the {\tt mov-to-cr3} instructions as
shown in Figure~\ref{fig:fastio_code}). These two instructions at
fastio entry/exit points are
the only occurrences of the 2.5-byte sequence in the guest's executable
address space.

The body of the fastio driver is our ``trusted computing base'' (TCB), as
it enjoys visibility into privileged state through the PPT. As discussed
earlier, we ensure that the TCB cannot be modified through EPT page-protection
bits. Further, we ensure that all execution within the TCB happens with interrupts
disabled (notice the {\tt cli} instruction at fastio entry), so that no other
code can run while the PPT is operational. We also
need to ensure that the fastio code is bug-free and cannot cause any exception,
lest the guest's untrusted exception handler may get called while the PPT is operational.
We also ensure that all non-maskable interrupts cause VM exits, so that the
hypervisor can interpose and disallow the guest from running while the PPT
is operational. Further, the TCB uses a separate CPU-private stack to disallow another
processor from trying to interfere with our execution by causing race conditions on
our stack state (lines 6,7,11 in Figure~\ref{fig:fastio_code}).
The CPU-private stack is also mapped in the PPT and the PGPA
space, to protect it from the rest of the guest kernel. Further, we ensure that
the pages belonging to the fastio driver are mapped correctly in the guest page
table (if mapped), and only in one place at its designated virtual address.

We ensure that the only {\tt mov-to-cr3} opcodes in the guest's executable address
space are
the ones belonging to the fastio driver at entry and exit.
The guest could potentially launch an attack, by directly jumping to one
of these two
{\tt mov-to-cr3} instructions inside the fastio driver, to subvert our
protection. We discuss three potential attack scenarios, and how we neutralize
them.

\subsubsection*{Jumping to one of fastio's {\tt mov-to-cr3} instruction with an
arbitrary value in the {\tt eax} register}
The guest could load a pointer to an arbitrary page table structure
in the {\tt eax} register and jump directly to one of the {\tt mov-to-cr3} instructions
in the fastio driver. This can cause an arbitrary page table to get loaded
within the guest. This new page
table could contain mappings to the PGPA pages; further,
the new page table may map the virtual address corresponding
to the EIP register to a totally different GPA, thus allowing execution of arbitrary
code while the PGPA pages are mapped.

Fortunately, this attack can be prevented by using a feature in x86 virtualization
hardware, called ``CR3 target controls''. Using this, it is possible to configure
the hardware such that VM exits occur on each execution of the {\tt mov-to-cr3}
instruction, except when the value being loaded into the {\tt cr3} register
is one of the values specified in CR3 target controls. The x86 hardware supports
specification of up to four target controls, and this capability was perhaps included
for efficient shadow-page-table based virtualization.

We use this interesting capability in the following way: we configure the hardware
to exit whenever the guest executes the {\tt mov-to-cr3} opcode, except when the
value being loaded in the {\tt cr3} register is the PPT address. We do so by specifying
the PPT address as one of the CR3 target controls. We use the other three CR3
target controls as a cache for recently seen values of the {\tt cr3} register
at the time of the call to the fastio driver.

Assuming that there are a small number of processes (typically one) accessing
the fastio driver, the cr3 values for these processes would get cached in the
CR3 target controls. Hence, VM exits would be avoided at fastio entry (because
the loaded CR3 value is the PPT address which is one of the CR3 target controls)
and at fastio exit (because the loaded CR3 value would typically be one of the
cached values in the CR3 target controls). In all our experiments involving
netmap, there was only one value of cr3 observed at fastio entry, which was
easily cached using the CR3 target controls, resulting in exitless
guest-fastio-guest path.

Now, the original attack, whereby the guest can jump to fastio's mov-to-cr3
instruction with an arbitrary value in {\tt eax}, is thwarted because the value
in the eax register is not going to be one of the CR3 target controls. Hence,
a VM exit would occur and the hypervisor can interpose and prevent the attack.

Further, the guest may try to set its {\tt eax} register to one of the
cached target controls, and then branch to fastio's mov-to-cr3, thus avoiding
the VM exit. The only security threat from this behaviour could occur if the
guest uses the PPT's address in the {\tt eax} register; all other cached
target controls do not map the PGPA space. We next discuss these attacks in
detail.

\subsubsection*{Jumping to fastio's entry {\tt mov-to-cr3} instruction with
PPT's address in the {\tt eax} register}
The guest could jump directly to the first {\tt mov-to-cr3} instruction
(at fastio driver entry) without disabling interrupts. Because the guest
will only try to load the PPT into the cr3 register, a VM exit will not
occur (as the PPT address is one of the CR3 target controls). This can potentially
allow the
guest to receive an external interrupt (as the interrupts were
not disabled), and execute its untrusted interrupt handler
while the PPT is operational. To prevent this attack, we configure
the virtualization hardware such that all external interrupts cause VM exits.
Because our experiments involve
high-throughput workloads, our fastio driver operates in polling
mode, and this extra interrupt handling cost does not cause
performance degradation. In Section~\ref{sec:discussion},
we discuss a solution which allows the guest to directly handle
hardware interrupts (through Intel's VT-d posted interrupts hardware
feature \cite{intel_vtd_manual}) without VM exits, and yet ensures VM exits
on any interrupts received while the PPT was operational.

Thus, we can effectively ensure that the hypervisor
gets to interpose on any external
interrupt received while the PPT was operational. The hypervisor
identifies the attack by determining if the PPT was operational while the
interrupt was received,
and thwarts it (potentially by terminating the guest).

\subsubsection*{Jumping to fastio's exit {\tt mov-to-cr3} instruction with
PPT's address in the {\tt eax} register}

The guest could load the PPT's address in a register and branch to the
second {\tt mov-to-cr3} instruction (at fastio driver exit). This can enable the
guest to execute untrusted code inside fastio's caller while the PPT is operational.
To thwart this attack, we further add a check at fastio's exit (after the {\tt mov-to-cr3}
instruction has executed) to confirm that the current value of {\tt cr3} is not the
PPT's address. If it is, the guest makes a hypercall to
alert the hypervisor. Because interrupts are not allowed while this code is executing, and
the PPT is guaranteed to map the fastio code pages correctly (and only in one place),
this ensures that such an attack gets thwarted.

If the guest tries to jump anywhere in the middle of our fastio driver, it does not
pose a security risk, as that cannot allow the guest to load the PPT or otherwise
obtain the capability to map the PGPA pages in its GVA space, other than in ways
discussed above.

\section{The fastio driver}
\label{sec:fastio}
The fastio driver is our privileged code (TCB) can access both guest's data structures
and the hypervisor's privileged state (including device state). The driver acts as a
bridge between the guest kernel and the device, and also allows sharing of the device
among multiple VMs and the host.
\subsection{Setup}
To simplify design, we use the same
fastio driver, both within the guests, and at the host. The host's fastio driver performs
a few extra operations related to initialization of the actual hardware device. At boot time,
the host loads the fastio driver, thus initializing the hardware device, and initializing
a PPT for its own use.
The host's PPT maps the host kernel (at their original virtual addresses)
and the device pages at a fixed virtual address, say PDVA (privileged device virtual
address). After host's PPT has been correctly initialized, the host can now use the
fastio driver to communicate to/from the device. For example, a transmit call from
the host involves switching to the PPT (within the host) and transferring packets
from the host kernel to the device. For 32-bit Linux, we use the top 1GB
of the VA space for the host kernel (0xc0000000-0xffffffff), and use
516 contiguous pages starting at PDVA to map the device state. For our prototype,
we use PDVA = 4MB (it must not overlap with the kernel's address space).
Of the 516 device pages in the PPT, 512 pages are for device MMIO, and four pages
are for storing the device rings.
Because we use the netmap API for the device driver, whereby all rings and buffers are
pre-allocated by the kernel, the fastio driver is also responsible for allocating the
host's netmap rings/buffers at load-time.

The initialization of the guest-side fastio driver is relatively simple. The fastio driver
authenticates itself to the host, and allocates its netmap rings/buffers, and ensures
that the guest kernel can see an attached NIC. The guest's fastio driver also initializes
its PPT, which contains mappings for the device state. These mappings are made at the
same virtual addresses in all guest/host PPTs.

Besides device state, we also need mappings inside the PPT
for the buffers and rings of the host and other guests. This is
required to allow sharing of the network device; one guest should be able to
receive packets for other guests/host inside its fastio driver.
We follow a convention, whereby
each guest (and host) is allocated a fixed amount of VA space in the PPT, to map its
network buffers. In our experiments, a 16MB space for each guest is enough to map all
its network rings and buffers. These mappings start at a fixed address,
called $ppt\_va\_start$, which must be distinct from the kernel addresses, and the
addresses used to map the device state. In our prototype implementation for 32-bit Linux,
we use {\tt ppt\_va\_start} = $16MB$. The device state is mapped below {\tt ppt\_va\_start}
and the kernel is mapped starting at address {\tt c0000000} (3GB). Thus, the VA space between
16MB and 3GB is available for mapping guest buffers and rings. Assuming 16MB per guest, this
allows us to map up to 191 guests inside the PPT at the same time.

Each guest's slab of VA space in the PPT is laid out in a fixed format: the first few
pages are dedicated to the transmit and receive netmap rings, and all the other
pages contain the guest's network buffers. A slot inside a netmap ring contains pointers
to the network buffers. Alongwith the original buffer pointers (pointing to the
guest kernel addresses
of the buffers), we also keep the corresponding
PPT pointers (pointing to the the same buffer but through the PPT addresses) with each netmap
ring slot. The PPT pointers are initialized at fastio load time, as discussed below.

To distinguish between guests, we assign a fastio ID to each guest/host. The host always
has ID 0, while guests are given IDs dynamically by the host, using bitmap-based allocation.
The netmap buffers of guest ID $n$, are accessible at
virtual address {\tt ppt\_va\_start + $n$ * 16MB} in the PPT. Each guest slab (of 16MB) contains
its netmap rings and netmap buffers, laid out sequentially and contiguously.

Mapping between the slabs at {\tt ppt\_va\_start} and the corresponding physical memory
need to be created dynamically, as guests boot and shutdown. The mapping for the host's slab
remains static, and gets initialized at PPT creation time (alongwith the mappings for the device
state). When the fastio driver initializes within the guest, it allocates its netmap rings
and buffers, and makes a hypercall to inform the host about the GPA addresses of these
rings/buffers. The host translates the GPA addresses to the corresponding host physical
addresses (HPA), and creates mappings appropriately in the host's PPT. 
Even if the buffers/rings
were discontiguous in the guest kernel, they are laid out sequentially and
contiguously in the PPT's
address space.
Further, the host walks through the netmap rings of the guest, and initializes the
PPT pointers inside the netmap ring slots. (Recall that
the PPT pointer for a guest buffer is
the address at which that buffer was mapped in the PPT in the guest's slab). These
pointers can now be
used by the host/other guests to access this guest's network buffers.

The PPT pointers stored in netmap ring slots are now visible to the untrusted guest, and
the guest could potentially modify these pointers to try and confuse the fastio driver.
To avoid this attack, our fastio driver performs a base-and-bounds check on the
PPT pointer before de-referencing it: ({\tt ppt\_pointer} $\geq$ {\tt ppt\_va\_start} + $n$ * 16MB)
and ({\tt ppt\_pointer} $<$ {\tt ppt\_va\_start} + $(n+1)$ * 16MB), where $n$ is
the guest ID.
This base-and-bounds check 
ensures that the PPT pointer lies within the guest's PPT slab, and so
the guest cannot cause the fastio driver to incorrectly read/write to memory outside
its own address space.

Because we have multiple PPTs (one for each guest, and one for the host), we may need
to create these mappings inside all PPTs. For the host, we create these mappings immediately
(at the time of fastio initialization hypercall). For the guests, we create these
mappings on-demand as follows:
\begin{enumerate}
\item All guest PPTs are initialized such that they contain GVA$\rightarrow$GPA
mappings from {\tt ppt\_va\_start} (in GVA space) to $pgpa\_start$ (in GPA space) for
a contiguous block of size ({\tt MAX\_GUESTS} * 16MB). $pgpa\_start$ is a
PGPA address, i.e., an address in the guest's physical address space which
is distinct from guest's physical memory. We use $pgpa\_start = 4GB + 16MB$.
(Recall that the PGPA addresses in our implementation start at 4GB and
are used to map privileged device and hypervisor state).

Thus, the mappings for guest ID $n$ are accessible at
address $pgpa\_start + n * 16MB$ in the GPA space (for all guests). Initially
(when there are no guests),
these GPA addresses do not map to any host physical addresses (HPA), i.e., the
present bit in the corresponding EPT entries is set to zero.
\item If a guest tries to access the netmap buffers of another
guest (or its own netmap buffers through the PPT),
an
EPT violation may result if the corresponding GPA address (in the {\tt pgpa\_start} region)
is currently unmapped.
If this happens, the hypervisor handles the EPT violation by creating
the required GPA$\rightarrow$HPA mapping on demand, before resuming the guest.
\end{enumerate}

\subsection{Transmit/Receive Path}
\label{sec:txrx_path}
The fastio driver is trusted and works cooperatively with all the other
fastio drivers to transmit/receive network packets.
Mutual exclusion among
different fastio drivers, is
ensured through a shared lock, which is also mapped using the PPT in
all guests/host.

On the transmit path,
the driver flushes its own tx buffers into the hardware
ring
On the receive path, the driver consumes packets from the hardware
ring, determines the destination for each packet, and copies that packet into
the rx buffers of the appropriate guest/host. (Recall that using PPT, guest ID,
and PPT pointers, any guest/host can read the rings/buffers of any other
guest).


On the receive path, it is possible for a guest to behave ``selfishly'' by
never calling into the fastio driver, and still enjoying the ``service'' of other
guests (as this guest's packets would still be received by other
guests). Such an attack can easily be prevented by maintaining statistics on the
number of fastio calls by each guest, and selectively dropping guest packets
on noticing selfish behaviour. Even if all guests are selfish, the host
would still be able to receive packets for all of them and for itself.


While we
use coarse-grained locking (one shared lock), finer-grained synchronization
could
potentially increase
concurrency and perhaps performance. Further, in high-contention
scenarios, it may be better to select a ``leader'' (or a few
leaders) which will be responsible
for switching packets for all guests/host.
Priorities could also be introduced during acquisition of the shared lock,
if needed.


\subsection{Zero-Copy}
Finally, we try and eliminate packet copies between the hardware ring and
the guest/host netmap rings. On the transmit path, zero-copy is
straight-forward. The fastio driver
maintains a hash table, which contains a mapping between
PPT pointers (for guest's network buffers) and their
corresponding HPA addresses. Insertions to this table
happens at fastio driver initialization time (when PPT pointers
are determined by the host during the hypercall). Further the host
pins these HPA addresses to memory, i.e. it ensures that its swapper
will never swap these addresses to disk.

During transmit, the fastio driver performs a fast hash lookup to convert the
PPT pointer of the network buffer (to be
transmitted) to its HPA address, and writes the computed HPA address
to the hardware ring, thus avoiding packet
copies. Using this mechanism, the hardware transmit
ring would now contain HPA pointers to buffers belonging to different guests
simultaneously.

On the receive path, zero-copy is similar. The guests/host provide PPT pointers to
empty buffers, through their netmap rings. The host converts the PPT pointers to
their HPA addresses and writes them to the hardware ring. Again, the
hardware receive ring would now contain HPA pointers to buffers belonging
to different guests simultaneously. The hardware stores the received packets
into these buffers. There are two possibilities on the receive path: either
the packet-destination is the same as the owner of the buffer in which it was
received (match); or the packet-destination is different from the buffer owner
(mismatch).

Matches are easy to handle --- we simply enqueue the buffer pointer to the
destination guest's netmap receive ring, so the guest can read the contents of the
received
packet. Mismatches cannot be handled in this way --- we cannot
enqueue the buffer pointer to the
destination guest's receive ring, as the buffer does not belong to
this guest. For mismatches, we allocate a fresh buffer from the destination
guest, and copy the packet contents into it before enqueueing the newly allocated
buffer into the destination guest's netmap receive ring.

There are two caveats to receive-side zero-copy. First, it is now possible for
one guest to snoop on the packets of another guest (if the first guest's buffer
is used to receive the second guest's packet). This opens the
possibility of one guest launching a man-in-the-middle attack on another guest.
Because modern network stacks are usually resilient to man-in-the-middle
attacks (through end-to-end encryption, for example), this is usually not an
issue. Second, the total number of buffers available to implement the receive
stack is now smaller: without zero-copy, there was an extra set of buffers
available exclusively to the hardware ring; with zero-copy, the hardware ring
relies on the buffers provided by the guest/host netmap rings. The extra set
of buffers available to the hardware ring, allow the ``double-buffering'' effect,
whereby the hardware can receive packets on its own set of buffers, while the
user application can read already received packets. To avoid this downside,
we allocate netmap rings with twice the number of buffers as the hardware ring.
This ensures that the double-buffering effect remains intact, even in
zero-copy mode.

Due to mismatches, the order of received buffers can be different from the
order in which buffers were allocated in the ring.
Because, zero-copy receive has slightly
weaker security guarantees, and requires more memory, we show results both
with and without rx-zerocopy.

\section{Experiments}
\label{sec:experiments}
We conducted our experiments on a 4-core machine with 16GB RAM and an Intel
X540-T2 10Gbps network adapter. The machine was connected to a 10Gbps network
switch. To send/receive packets at the other end, we used another machine with
an identical 10Gbps network adapter, also connected to our 10Gbps network
switch. We used 32-bit Linux 3.9.0 on our hosts and guest with
PAE-paging mode. Our guest was configured
with two CPUs, and 1GB memory. For experiments involving a single-core host,
our guest was also given only a single CPU.
For our netmap-based experiments, we used netmap's pkt-gen utility to
send/receive packets.
For experiments involving socket-based I/O, we used the netperf
utility.

Our fastio driver is based on the Linux ixgbe driver, with the netmap ixgbe
patch. The netmap's patch defines two functions, ``txsync'' and ``rxsync'', which
are used to transmit/receive packets between the user's netmap rings and the
hardware device respectively. The transfers between the netmap rings and the
hardware devices are performed in batches, where the batch-size depends on the size
of the hardware ring. We used a hardware ring with 512 slots (default).
To implement fastio, we modified the txsync and rxsync function to switch to our
PPT at entry, and switch back to the original page table at exit. i.e.,
netmap's txsync and rxsync functions form the body of our fastio driver
(Figure~\ref{fig:fastio_code}). We configured the driver to use polling, to
avoid extraneous scheduling issues during our experiments \cite{receive_livelock}.
Our source code and raw data will be made publicly available.

Because the txsync and rxsync functions execute within our
trusted fastio driver while the PPT is operational, they
can access the rings of all guests/host, as well as the hardware ring using the
PPT addresses. We implemented our zero-copy logic within the txsync and rxsync
functions. The same logic (for txsync and rxsync) would execute both within the
host and within all guests, as also discussed in Section~\ref{sec:txrx_path}.

Table~\ref{tab:thruput} presents our results for the
network throughput for a single guest on a multi-core host. 
The rows labeled {\tt netmap-} use the netmap API for user/kernel
communication within the
guest, while the rows labeled {\tt socket-} use the socket API.
We show results using the virtio interface without/with vhost support
(labeled {\tt -virtio-} and {\tt -virtio-host-} respectively).
Further, we show results for two different types of host-side switches,
namely {\tt tap}, and {\tt netmap}. For the netmap ``switch'', the netmap
ethernet interface was directly exposed to the Qemu process (without using
the VALE switch that ships with netmap). The performance with the VALE
switch is inferior to the performance without it, as it
adds extra computation on the switching path.

Without the netmap API within the guest, the throughput is heavily
CPU-bottlenecked at the guest's user/kernel interface. The netmap API 
on baremetal host, is able to saturate the 10Gbps NIC even
with 60B packets (14.8 Mpps). However using netmap with virtio incurs
a
large performance penalty (2.3 Kpps for 60B packets). virtio-vhost-tap
improves the throughput marginally, while virtio-netmap does not perform
any better. We were unable to setup {\tt virtio-netmap} to receive
packets; the transmit side throughput in this configuration is
only 154Kpps, and we do not expect the receive throughput to be
significantly higher.

These CPU bottlenecks have also been previously reported
by ClickOS \cite{clickos}. The ClickOS project addresses this problem
by overhauling Xen hypervisor's I/O virtualization subsystem. Just like
netmap, ClickOS optimizations involve memory-preallocation, batching,
and fast switching, albeit at the VM/hypervisor interface. Even with
all these optimizations, ClickOS peaks at around 11 Mpps while transmitting
60B packets using a 512-slot device ring. On the receive side, ClickOS
peaks at around 6.2 Mpps for 60B packets using a 512-slot device ring.
(These figures have been taken from \cite{clickos}).

The row labeled {\tt -fastio} shows
the throughputs achieved by our solution. Expectedly,
our achieved throughputs are
very close to the throughputs achieved on bare-metal. Compared to ClickOS,
our transmit throughput is around 33\% better, and our receive throughput
is around 115\% better, for the same hardware configuration. The row
labeled {\tt -fastio-no-rzc} shows throughput without our
zero-copy optimization on the receive side; because the cost of packet-copies
is not significant on the I/O path, the throughputs are largely similar to
{\tt -fastio}.

Table~\ref{tab:thruput_uni} presents the throughput results for a uniprocessor
host and guest. Our solution remains unaffected by the scarcity of CPUs, while
virtio and virtio-vhost observe (sometimes significant) performance penalties.

Table~\ref{tab:multiguest} shows the total throughput with multiple transmitters
and receivers for 60B/1500B packets. For multiple tx/rx agents ($>1$),
one of the agents is the host and the rest are VMs. We show results
for up to four tx/rx agents, as our test machine had four CPU cores.
The transmit-side throughput remains largely unaffected with increasing
number of transmitters; moreover, the bandwidth is largely fairly allocated
among the transmitters. On the receive-side, we notice throughput degradation
for small packets with increasing number of agents; this degradation is largely
due to packets dropped while trying to enqueue them to the receive-side rings
of other agents. For a single receiver, packet drops cannot not happen, as
the receiving agent simply returns to its userspace (which consumes the
packet) on observing a full ring. However, if one agent tries to
enqueue a packet to a full ring of another receiver, the packet gets
dropped (wasted work). The probability of packet drops increases
with increasing number
of receivers. The probability is smaller for larger packets, as the CPU
has to do less work --- thus we do not see the effect of dropped packets
for 1500B-sized transfers.

\begin{table}
\begin{center}
\begin{tabular}{c|c|c|c|c}
 Number of tx/rx & 1 & 2 & 3 & 4\\
\hline
 tx-60B & 14704 & 14753 & 14776 & 14860\\
\hline
 tx-1500B & 815 & 820 & 820 & 820\\
\hline
 rx-60B & 13292 & 11712 & 9800 & 8311\\
\hline
 rx-1500B & 816 & 820 & 820 & 820\\
\end{tabular}
\caption{\label{tab:multiguest} Transmit and receive performance for multiple VMs on a 10Gbps NIC (Kpps).}
\end{center}
\end{table}

Next, we discuss the maximum achievable throughput of our fastio
packet switch, assuming both the transmitter and receiver are running
as VMs (without NIC involvement).
For this experiment, we implemented a shared software ring which replaced our hardware
ring. The transmitters enqueued to this software ring, while the receivers dequeued
from it. For this experiment, the transmit-side operation involved copying
packets (always)
from netmap rings to the shared software ring; on the receive-side, we show results
with and without the zero-copy optimization.
Table~\ref{tab:switch}
presents throughput results for one-transmitter/one-receiver (tx1-rx1),
one-transmitter/three-receivers (tx1-rx3), three-transmitters/one-receiver (tx3-rx1),
and two-transmitters/one-receiver (tx2-rx2). The maximum achievable throughput with
receive-side zero-copy enabled, is around 30 Mpps for small packets, and 10.9 Mpps for
large packets; without zero-copy, the throughputs decreases to 22.6 and 7.8 Mpps
respectively. In all these cases, the achieved throughputs are well above
the line-rates supported by current NICs. With increasing number of receivers, packets
start getting dropped resulting in lower overall throughput. With an increasing
number of transmitters (tx3-rx1), the throughput drops marginally, presumably due to
lock contention.
VALE, another netmap-based software switch, reports throughputs of
around 3.4 Mpps (tx) and 2.5 Mpps (rx) while running KVM-based virtual machines
\cite{vale}.
While exact/fair comparisons with VALE are not possible (as our
switch is perhaps lacking in many features provided by VALE), the performance
improvements provided by our switch due to in-guest switching are clearly
visible. In contrast, VALE requires host-side involvement.
\begin{table}
\begin{center}
\begin{tabular}{c|c|c|c|c}
     Config & \multicolumn{2}{c|}{fastio} & \multicolumn{2}{c}{fastio-no-rzc}\\
    \hline
     pktsize & 60B & 1500B & 60B & 1500B \\
    \hline
     tx1-rx1 & 30.43 & 10.93 & 22.59 & 7.84 \\
     tx1-rx3 & 12.08 & 6.18 & 11.37 & 5.21 \\
     tx3-rx1 & 27.19 & 8.02 & 20.06 & 6.20 \\
     tx2-rx2 & 15.61 & 7.50 & 13.80 & 5.81 \\
\end{tabular}
\caption{\label{tab:switch} Total throughput for software-only switching without NIC involvement (Mpps).}
\end{center}
\end{table}

Finally, we discuss the runtime overheads of DBOS. DBOS overheads are related
to VM exits caused due to guest's execution of the {\tt mov-to-cr3} opcode (which
is patched by us to the {\tt int3} opcode), and VM exits due to write-accesses to the
page-table pages (which are write-protected by us using the EPT).
For all our experiments involving
{\tt pkt-gen} and the Linux kernel, we encountered zero overhead due to DBOS. This is
expected because only a few page-table switches occur during {\tt pkt-gen}
execution, and almost no writes happen to the page-table pages. However, it is possible
for a VM to be running other programs simultaneously with {\tt pkt-gen}; to characterize
these overheads, we run some CPU-intensive programs (taken from SPEC
CPUInt2000 \cite{spec:cint2000})
and present runtime overheads, alongwith the statistics collected for VM exits. The
overheads range between -2\% and 34\%; the majority of this overhead
is due to writes to page-table pages by the guest kernel, presumably to implement
LRU page-replacement algorithm. We also show results for the {\tt forkwait} microbenchmark
\cite{adams:asplos06} which forks 40,000 processes and waits for each of them to
exit in turn. Given that the forkwait benchmark
creates and destroys a large number of page tables, the resulting DBOS overhead is
significant (18.74x). After running all these benchmarks, 12 kernel pages contained
at least one {\tt int3} patch.
All these patches were in the guest's kernel, and were due to
{\tt mov-to-cr3} instructions. We also found 193 instances where the suffix of the
2.5-byte sequence occurred at the top of an executable page in the guest. Of these, 39
occurred in kernel's executable pages and 154 occurred in user's executable pages.
Similarly, we found 69 instances where the prefix of the 2.5-byte sequence occurred at
the bottom of an executable page in the guest (60 user, 9 kernel).

\begin{table}
\begin{center}
\begin{tabular}{c|c|c|c|c}
             & original & DBOS     & & \\
     program & time (s) & slowdown & ptable-exits & cr3-exits\\
     \hline
     gcc & 35.6 & {\bf 1.34} & 910509  & 2436\\
     perlbmk & 6.2 & {\bf 1.13} & 54515 & 619\\
     gap & 46.7 & {\bf 1.03} & 113588 & 1593\\
     bzip2 & 103.8 & {\bf 1.05} & 420005 & 3783\\
     twolf & 119.1 & {\bf 1.00} & 22330 & 3906\\
     gzip & 94.4 & {\bf 1.08} & 560305 & 3820\\
     vpr & 79.2 & {\bf 0.98} & 120257 & 2659\\
     mcf & 40.1 & {\bf 1.02} & 56913 & 1372\\
     crafty & 45.3 & {\bf 1.01} & 14418 & 1545\\
     forkwait & 8.9 & {\bf 18.74} & 10924572 & 166187\\
\end{tabular}
\caption{\label{tab:spec} DBOS overheads}
\end{center}
\end{table}

\section{Discussion}
\label{sec:discussion}
Our performance experiments and our discussion on security demonstrates the utility
of DBOS as a security technique. We use several x86 mechanisms to achieve a practical
implementation of DBOS, namely, EPT-based read/write/execute page-protection,
the length of the {\tt mov-to-cr3} instruction opcode, support for CR3 target controls,
ability to configure the virtualization hardware to cause VM exits on certain
instructions, to name a few. For example, if the length of the {\tt mov-to-cr3} opcode
was smaller (e.g., one byte instead of the 2.5 bytes), the number of patch-sites
and resulting exits would have been greater. Similarly, security would not have been
possible without support for CR3 target controls.


As we discuss in Section~\ref{sec:security}, we configure the
virtualization hardware to cause a VM-exit on every interrupt, and
we discussed why we need this capability to ensure security (Section~\ref{sec:security}).
For interrupt-intensive workloads, this may be a severe performance
penalty \cite{eli}. Recent support for x86 VT-d posted interrupts allow the guest to directly
receive interrupts without requiring VM exits. Even if the guest wants to use VT-d posted
interrupts, we could still ensure that any interrupt received within the TCB causes a VM
exit by ensuring that the virtual addresses containing the interrupt-descriptor-table (IDT)
are unmapped in our PPT. Typically, the kernel initializes the IDT in
the beginning and stores its virtual address
in the IDTR using the {\tt lidt} instruction. The hypervisor can interpose on the
execution of the {\tt lidt} instruction (by requiring
a VM exit), and record the address of the IDT. Thereafter, it
can ensure that the virtual addresses corresponding to the IDT addresses in the PPT
are mapped to a {\em shadow} IDT \cite{eli}. All entries in our shadow IDT would have their
present-bit set to 0, causing a not-present exception an an interrupt. Additionally,
the host is configured to force a VM exit whenever a not-present exception occurs.
Through this, the hypervisor would get to interpose on any
interrupt received during TCB execution.

Because we ensure that all {\tt mov-to-cr3} instruction executions within the guest
cause VM exits, this can cause overhead for applications that involve significant
context-switching, as also seen in our experiments. One could potentially optimize
this further by using in-place binary patching to replace all {\tt mov-to-cr3} instructions
with a call to a special trusted function (another TCB), that allows loading the cr3
without requiring a VM exit. This would involve caching the most-frequently-used cr3
values in the CR3 target controls. We leave this optimization for future work.

The choice of the opcode to subtract is also interesting; on the x86
architecture,
we identified three different possible ways of accomplishing security
through opcode subtraction. We have discussed the first one
involving subtraction of the {\tt mov-to-cr3} opcode
in this paper. The other two involve subtracting either the {\tt lgdt} opcode,
or the {\tt mov-to-cr4} opcode. In both these cases, the subtracted opcode is used
at entry and exit of the fastio driver. Of all the three choices, the {\tt mov-to-cr3}
opcode incurs the least overhead on the I/O path.


Finally, we discuss guest fidelity. In general,
DBOS does not change guest behaviour in any
way (apart from potentially slowing it down). The only exception is that we
rely on the identification of instruction boundaries; if instruction
boundaries are incorrectly determined, or they can change dynamically, DBOS
can cause the guest's logical behaviour to change. As we show in our experiments
for Linux, this is usually not an issue. While this work is about implementing
DBOS on existing OS/programs, it should be possible to make the compiler
DBOS-aware, so that it prevents emission of certain byte sequences.

\section{Related Work}
\label{sec:relwork}
There are two categories of related work for this paper: one involving
network I/O optimization and virtualization, and the other involving software-based
security techniques such
as dynamic binary translation, proof-carrying code, and
typed-assembly language.

\subsubsection*{Network I/O Optimization and Virtualization}
Routebricks \cite{dobrescu09} worked on implementing
fast software routers by scaling them on
a number of servers. PFQ \cite{bonelli12}, PF\_RING \cite{pf_ring},
Intel DPDK \cite{dpdk}, and netmap \cite{netmap} are all approaches
involving mapping NIC buffers into user address space.

As already discussed,
ClickOS \cite{clickos} suggests a complete overhaul of the Xen hypervisor's
network interface, which increases the effective bandwidth between the
guest OS
and the hardware NIC.
Other work on improving hypervisor networking performance \cite{ram09,
santos08, cardigliano11, rizzo13} suggest similar optimizations;
\cite{xu13, nadav13}
discuss scheduling optimizations for good networking throughput.
Efforts involving vhost-like optimizations for
Open vswitch \cite{openvswitch, ram09} are also interesting.
All these efforts involve optimizing either the guest-side stack, or
the host-side stack, or the guest/host virtualization interface.
In comparison, our approach completely obviates the host-side
stack, and provides direct NIC access to the guest, resulting in
significantly higher throughputs and lower latencies.

\subsubsection*{Software Techniques for Security}
The closest competing technique to DBOS, is perhaps dynamic binary
translation (DBT). Unlike DBOS, DBT incurs large overheads
for indirect jumps and interrupts/exceptions \cite{adams:asplos06, drk}.
BTKernel \cite{btkernel}
optimizes DBT for interrupts/exceptions and indirect
branches; however, BTKernel cannot provide the security guarantees
required for our application. For example, BTKernel's
approach of leaving code-cache address on return stacks, and
jumping directly to them can be used to launch a security
attack in our case.
DBOS is a low-overhead mechanism for
ensuring security, and usually results in much lower overheads
than DBT for similar security guarantees.
Conversely, DBOS is
not as powerful as DBT, and cannot be used for several other DBT
applications.

DBOS is similar to
verification techniques such as proof-carrying code (PCC) \cite{pcc} and
typed-assembly language (TAL) \cite{tal}, in that, both
techniques involve analyzing the code at load time to ascertain
safety. However, unlike PCC and TAL, our analysis is much
simpler --- we only check for the occurrence of a certain
pattern (grep) in the executable address space. In contrast,
PCC and TAL require detailed reasoning about semantics of individual
instructions, and control flow. While PCC and TAL have been successfully
used to ascertain safety for relatively small programs, ascertaining
safety against a full guest operating system still remains an open
problem. Also, verification techniques seldom worry about instruction
boundaries, and the potential of being able to jump in the middle of
an instruction. Our current method for driver certification
involves digital signatures; it remains to be seen if methods like
proof-carrying code may be used instead. A
PCC-based certifier would need
to certify that the fastio driver behaves as expected, and does
not allow PPT access to be leaked to the untrusted guest.

\section{Conclusions}
\label{sec:conclusion}
We present a novel security mechanism, DBOS, and show its successful
application for I/O virtualization. Using DBOS, we are able to expose
the privileged hypervisor/device state to the guest without
security risks. Our trusted guest-side driver can access this
privileged state to perform fast I/O and switching. We show
significant improvements over the state-of-the-art device
virtualization solutions, and software switches.

\bibliographystyle{acm}
\bibliography{fastnet}

\end{document}